\documentclass[preprint,aps,showpacs]{revtex4}
\usepackage{graphicx}
\def\pb{$^{208}$Pb}
\def\be{\begin{equation}}
\def\ee{\end{equation}}
\begin{document}
 
\title{Fitting theories of nuclear binding energies}
\author{G.F.~Bertsch, B.~Sabbey, and M. Uusn\"akki}
\affiliation{Physics Department and Institute for Nuclear Theory, 
University of Washington, Seattle WA }
\date{\today}
 
\begin{abstract}

In developing theories of nuclear binding energy such as
density-functional theory, the effort required to make a fit can
be daunting due to the large number of parameters that may be in
the theory and the large number of nuclei in the mass table. 
For theories based on the Skyrme interaction, the effort can be
reduced considerably by using the singular value decomposition to
reduce the size of the parameter space.  We find that the
sensitive parameters define a space of dimension four or so, and
within this space a linear refit is adequate for a number of
Skyrme parameters sets from the literature.  We do not find
marked differences in the quality of the fit between the SLy4,
the BSk4 and SkP parameter sets.  The r.m.s. residual error in
even-even nuclei is about 1.5 MeV, half the value of the liquid
drop model.  We also discuss an alternative norm for evaluating
mass fits, the Chebyshev norm.  It focuses attention on the
cases with the largest discrepancies between theory and
experiment.  We show how it works with the liquid drop model and
make some applications to models based on Skyrme energy
functionals.  The Chebyshev norm seems to be more sensitive to
new experimental data than the root-mean-square norm.  The
method also has the advantage that candidate improvements to the
theories can be assessed with computations on smaller sets of
nuclei.\\
\end{abstract}
 
\maketitle
\section{Introduction}

In making theories of nuclear binding energies (``mass
formulas"), there are invariably parameters that are determined
by fitting the experimental data.  If the fitting parameters
are not defined properly, they may be underconstrained by the data,
causing problems in making the fit.  This is the situation for
the Skyrme parameterization of the self-consistent mean-field
theory (SCMF) of binding energies, also  called  
density-functional theory.  While methods for dealing
with underconstrained parameters are well known, to our knowledge there has not
been a critical analysis of the Skyrme parameterization.  We shall
show here that the sensitive degrees of freedom in the Skyrme energy
functional are very similar to those of the liquid drop model, except for
the spin-orbit interaction.  When
the parameter space is restricted to those degrees of freedom,
linear methods can be applied to make a considerable reduction in 
the effort required for a search.  We show how this works in the
next section, refitting various parameterizations
from the literature.

A second way to reduce the effort in evaluating theories is to use
a more economical norm for the fit.  Usually one makes a least-squares
fit to the binding energies, i.e. using as a norm the 
root-mean-square (r.m.s.) average of the residual difference $r_A$ between 
theory and experiment, 
$$
r_A = E_{theory}(A) - E_{exp}(A).
$$
This requires calculating all nuclei.  An alternative is the
the Chebyshev norm, used in making a {\it
minimax fit}.   It allows one to screen
theories and their improvements taking only the information on 
a small set of nuclei, the ``worst cases" of the baseline theory.
By calling attention to these cases, it could also be helpful
to experimentalists choosing which nuclei to study, and to theorists
searching for missing ingredients in the baseline theory.
There is an extensive literature on the minimax method \cite{ch82},  
but it only rarely used in physics \cite{ja83}.  We describe the
method in Sect. III below, applying it to the liquid drop model as
a simple exercise and then to SCMF.
We also mention that
another norm
somewhat between the least squares and the C-norm has been proposed for
fitting nuclear binding energies \cite{audi2}. 

In our discussion below, the Bethe-Weizs\"acker semiempirical 
mass formula will provide a 
convenient calibration point on the theory.  For reference, the
formula is given by
\cite{po95}
\be
\label{ld}
B = a_v A - a_s A^{2/3} - a_c {Z^2\over A^{1/3}} - a_a {(N-Z)^2
\over 4 A} + \delta{ ( (-1)^N + (-1)^Z)\over A^{1/2} },
\ee
where $B$ is the (positive) binding energy of the nucleus $A$.

\section{Skyrme functionals}

The Skyrme energy functional in its present form \cite{be03} has 10 linear
parameters and one nonlinear parameter, the exponent $\alpha$ of
the density-dependent interaction.  We shall only fit the linear
coefficients, holding $\alpha$ fixed.  In practice, 
the spin-orbit interaction is often assumed to have a specific 
isospin structure.  This reduces the number of linear parameters to
nine.
Pairing is an indispensable ingredient in the
theory but is outside the 
scope of the Skyrme energy functional. We include pairing
following the treatment  of ref. \cite{be04}, using a Lipkin-Nogami
treatment of BCS pairing with a surface-enhanced zero-range pairing
interaction.  The fits are only done on even-even nuclei,
so the dependence of the fit on the binding energies would be
very much reduced anyway.
We represent Skyrme energy functional as in ref. \cite{be03} with
interaction energy density given by
\def\calI{{\cal I}}
$$
{\cal V} = \sum_{i=1}^{10} c_i f_i
$$
where $c_i$ is a parameter and $f_i$ is a function of the one-body
densities.  The specific forms of the $f_i$ needed for time-reversal
invariant densities are given in Table I.
\begin{table}      
\label{skyrme-def}
\caption{Skyrme energy functional in the $c$-parameterization.
The label
$i$ is expanded into a integer $n$ and an isospin label
$t=0,1$.  The isoscalar and isovector densities are $\rho_{0,1} =
\rho_n \pm \rho_p$ and other densities are the same as defined
in ref. \cite{be03}. }
\begin{tabular}{c|l}
\hline
$i=nt$ &    density \\
\hline
$1t$ & $\rho_t^2$  \\
$2t$ & $\rho_t^2 (\rho_0)^\alpha $ \\
$3t$ & $|\nabla \rho_t|^2$  \\
$4t$ & $\rho_t \tau_t$  \\
$5t$ & $ \nabla \rho_t \cdot {\bf J}_t$ \\
\hline
\end{tabular}
\end{table}
The interaction energy for each nucleus $A$ requires the integrals
over the corresponding densities $f_{iA}$, 
$$
I_{iA} = \int d^3r  f_{iA}.
$$

Since binding energy is a nonlinear function of the Skyrme parameters,
all of the complications of nonlinear searches are present in the
problem of optimizing the functional.   However, there are many 
parameter sets in the literature that we can take as starting
points in a search using linear methods.  This will determine a 
local minimum of the chosen norm in the vicinity of the starting parameter
set. On a technical level, one needs the derivatives of the binding
energies with respect to the parameters to make a linear fit.
The required derivatives are easily obtained with
the help of the Feynman-Hellman theorem \cite{FH} which expresses the derivatives
in terms of integrals over the original densities.
The theorem is valid for density functional theory because it has
a variational character.  The linear least squares fit requires inverting
the matrix $M$ having elements
\be
\label{M_mat}
M_{ij}  = (I I^\dagger)_{ij}  \equiv \sum_A I_{iA} I_{jA}.
\ee
The refit parameters are changed by
an amount $\Delta c$ given by
$$
\Delta c = M^{-1} I^\dagger r.
$$
Here $\Delta c$ and $r$ are vectors with components $c_i$
and $r_A$ respectively.
 
The main technical problem arises from the redundancy among the
parameters.  In parameter space, there are flat directions in which
the parameter change can be large with only a small
effect on the linearized energy.  However, if one accepts these
large changes, one discovers that they invalidate the linear
approximation and the final energies can turn out to be worse
than the starting values.  This is a common problem in fitting
and is typically addressed by a singular value decomposition of
the parameter space.  In essence, one reduces the dimension of
the space to eliminate the flat directions.  This is achieved
by first diagonalizing the matrix $M$ and projecting onto the
subspace spanned by the eigenvectors having eigenvalues 
larger than some minimum value.

We now show how this works in practice with the SLy4 Skyrme energy 
functional \cite{ch98}.  The binding energy calculations
are performed with the Paris-Brussels code
{\tt ev8} \cite{bo85} which requires that 
occupation probabilities of time-reversed orbitals are equal.
Accordingly, we only fit to the even-even nuclei in 
the mass table.  We also restrict the fit to nuclei with $N,Z \ge 8$, 
as the lightest nuclei are not well described by SCMF.  For the SLy4
interaction, the authors of ref. \cite{be04} provided us with
data on wave functions and equilibrium deformations that we took
as a starting point for our refits.
% Computer programs and files:
% Run ev8 on all nuclei with command line:   
%    ``python /scratch/refit/python/run.py''
% specify output directory and skyrme parameters in run.py using
%    parameters ``outdir'' and ``params''
%    for initial run without a parameter change (delta_C) set params=getSly4().
% Results for SLy4 are in /scratch/refit/sly4/sly4_results

\begin{figure}
\includegraphics [width = 11cm]{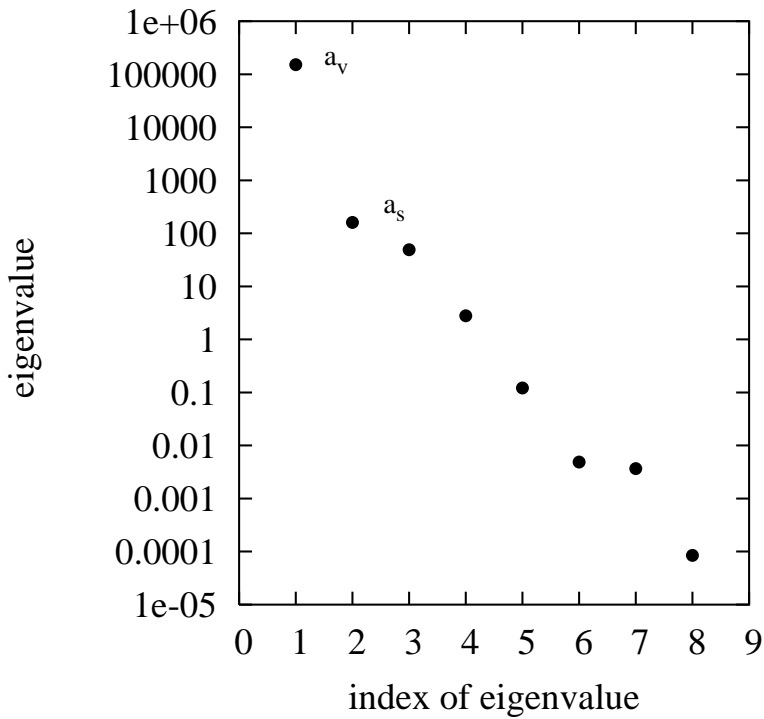}
\caption{Eigenvalues of the least square fitting matrix, eq. (\ref{M_mat})
}
\label{M_eig}
%fig 1 see /scratch/refit/plot/eigen2.gnu
\end{figure}

The experimental energies are taken from the 2003 mass table \cite{au03},
which has measured values for 579 even-even nuclei.
%  experimental binding tables in /scratch/refit/binding_tables
% mass.mas03    binding energy file from Audi 2003 see ref. [8]
% mass1995.dat                           Audi 1995
%read_audi2003.py  -->  exp.tab
% read_audi2003_200.py --> exp.200.tab
% read_audi1995.py     --> exp1995.200.tab
The SLy4 parameterization was obtained by constraining the parameters by properties
other than binding energy.  In particular, the equilibrium nuclear
matter density $\rho_{0,eq}$ was constrained to a value that yielded
a good description of nuclear radii. The condition may be 
expressed
$$
 {d \over d \rho_0 }{{ T +V}\over \rho_0} = 
{1 \over 5 m} {k_f^2\over \rho_{0,eq}}+ c_{10}
\rho_{0,eq} + (1+\alpha)c_{20} \rho_{0,eq}^{1+\alpha} + c_{40} k_f^2  =0. 
$$  
where $k_f= (3 \pi^2 \rho_{0,eq}/2)^{1/3}$ is the Fermi momentum.  
We enforce this condition in our fit,
leaving 8 independent parameters. We expect that the
resulting refit parameterization will maintain a good description of
nuclear radii.  
% fitting program:
%     cd /scratch/refit/python/distribute
%     python process_skyr.py /scratch/refit/sly4/sly4_results > sly4.dat
%     python cheby.py sly4.dat
% Output will contain eigenvalues of ``M'', experimental energies,
%    theoretical energies,
%    and minimax/least-squares refits for all numbers of vectors
The eigenvalues for the 8-dimensional matrix $M$ are displayed in 
Fig. \ref{M_eig},
ordered by size.  Remarkably, they span a range of
nearly 10 orders of magnitude.  Also note that the first
eigenvalue is 3 orders of magnitude larger than any of the
others.

We now consider fits in subspaces of various dimensions $N$, keeping
the $N$ most significant eigenvectors. 
The r.m.s. residuals of the linear
refits, 
$$
|r - (\Delta c I)^\dagger|^2 \equiv \sum_A ( r_A - \sum_i \Delta c_i I_{iA})^2,
$$
are shown in Fig.~2 as a function of $N$. The original SLy4 has a 3.3 MeV
r.m.s. residual, and this improves to about 1.7 MeV for $N\ge 3$.
Thus, it seems that spaces of dimension larger than 3 or 4 are not needed
to improve the quality of the fit.  However, since this is only a linear
refit it must be demonstrated that the quality of the fit
is not degraded when one uses the wave functions of 
the new parameter set.
\begin{figure}
\includegraphics [width = 11cm]{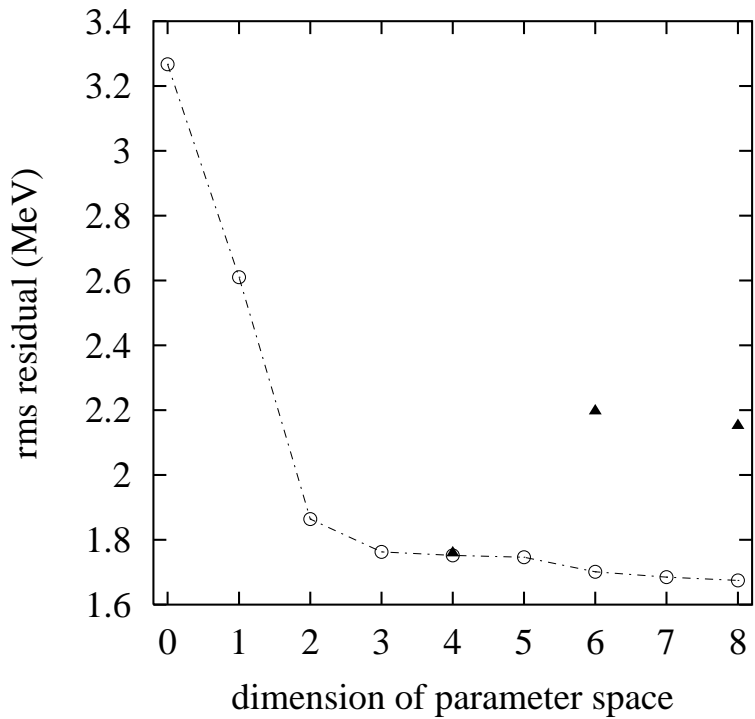}
\caption{Least squares refit as a function of the dimension of the parameter
space.  The r.m.s residuals of the linear refit are shown as circles,
connected by lines to guide the eye.  The triangles show the actual
r.m.s residuals for the parameters determined by the linear refit.
However, not all nuclei converged for the N=6 and N=8 parameter sets,
and they were not included in the r.m.s. residual.
}
\label{converg}
%fig 2  /scratch/refit/plots/converg2.gnu
% the refit data sets are /scratch/refit/sly4/sly4_refit_?vector  ?=4,6,8
\end{figure}
For the cases $N=4,6$ and 8, we have recalculated the orbitals of all the
nuclei to avoid the linear approximation.
The iterative procedure to recalculate the orbitals converged as 
expected for the $N=4$ parameter
space, and the new energies were very close to the values obtained
by the linear approximation.  For the
$N=6$ and $N=8$ spaces, there are large changes in some of the
parameters and the iterative procedure for calculating new
orbitals did not always converge.  The calculated norms including
only converged nuclei are shown as black triangles in Fig.~2.
One sees that the norm for the $N=8$ is much larger than
predicted by the linear approximation, and even in the $N=6$ case
there is significant error.  We show in Fig. \ref{th-exp} plots
of the residuals as function of $N$ for the SLy4 parameter set
and the $N=4$ linear refit.  One sees that the refit mainly
affects the heavy nuclei, correcting the trend to underbind
them.  This is at the expense of the region near the doubly magic
$^{208}$Pb, which is now overbound.
\begin{figure}
\includegraphics [width = 13cm]{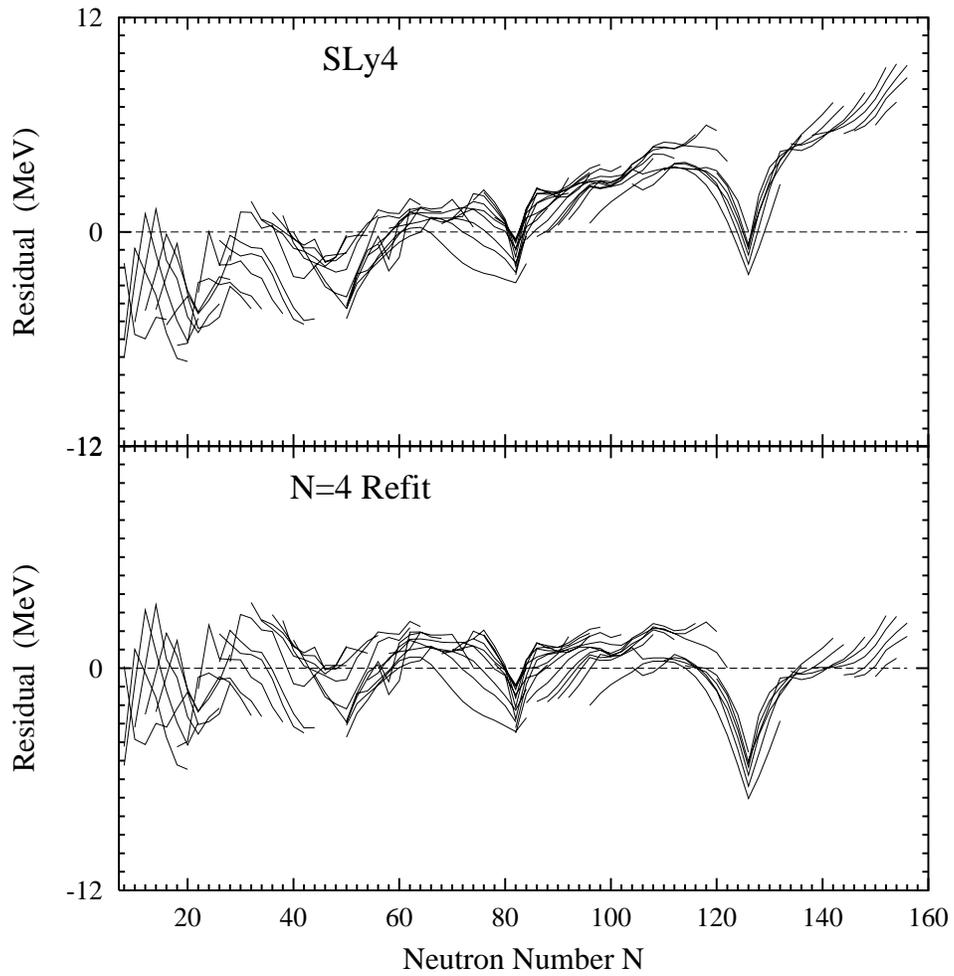}
\caption{Residuals between theory and experiment as a function of 
$N$, with nuclei of the same atomic number joined by lines.  The
top plot shows the original SLy4 and the $N=4$ linear refit is
shown on the bottom.
}
\label{th-exp}
% sly4 residuals are from process_skyr.py,cheby.py with 
% directory /scratch/refit/sly4/sly4_results
% full calculation linear refit is obtained from       
% process_skyr.py,cheby.py with directory
% /scratch/refit/sly4_results/sly4_refit_4vector
\end{figure}

Let us now try to interpret the eigenvectors of $M$.  The vector
corresponding to the largest eigenvalue is easy to interpret as
the
direction in parameter space that controls nuclear 
matter binding energy.  To see this, we start with the expression 
for the nuclear matter binding energy in term of the $c$ parameters,
$$
a_v(c) = { 3 k_f^2\over 10 m} + c_{10}\rho_{0,eq} +c_{20}
\rho_{0,eq}^{1+\alpha} + {3\over 5} c_{40} k_f^2 \rho_{0,eq}.  
$$
The direction vector with the maximum sensitivity to $a_v(c)$
is proportional to the gradient
$ \vec \nabla_c a_v(c)$.  We find that the overlap of this direction
vector with the first eigenvector of $M$ is 99\%.  Thus, the
extremely sensitive direction revealed by the eigenvalue plot
Fig. 2 is nothing more than the nuclear matter binding energy. 
Its new value in the refit is -16.06  MeV. 

Another important property of nuclear matter is the 
symmetry energy, given by
$$
a_s(c) = {k_f^2\over  6 m}  + c_{11}\rho_{0,eq} +c_{21} \rho_{0,eq}^{1+\alpha} + 
(c_{41} + {c_{40}\over 3} )k_f^2 \rho_{0,eq}.
$$
It does not correspond quite so well to a particular eigenvector:
the direction vector along  $\vec\nabla_c a_s(c)$ has a  53\% overlap 
with the
second and an 80\% overlap with
the third eigenvector.  However, within the space of the first three
eigenvectors, a vector with an overlap of 98\% can be
constructed. We can therefore say that the $N=3$ space includes parameters
for both the nuclear matter binding energy and the symmetry energy.
The three other adjustable parameters in the
liquid drop model are associated with the surface energy, the
Coulomb energy, and the pairing energy.  It is likely, but we
have not checked, that the surface energy of semi-infinite nuclear
matter provides a third vector for spanning the $N=3$ space of
the first three eigenvectors of $M$.  The remaining liquid drop
parameters are not relevant to our fits, since the Coulomb
interaction and the pairing are not adjusted.  The fourth
eigenvector of $M$ contains physics that is entirely missing
from the liquid drop model, namely the spin-orbit interaction. 
In fact, there is a 90\% overlap between that vector and the
relevant Skyrme parameter, $c_{5t}$ in the notation of Table I. 
Beyond the spin-orbit interaction, the Skyrme parameters do not
provide additional sensitive determinants of the binding
energies.

\begin{table}
\label{refit-ls}
\caption{Refits of various Skyrme parameterizations.  The
experimental data set is even-even nuclei of the 2003 mass table
\cite{au03}.
The last entry give the corresponding
properties of the liquid drop model, eq. (1), for comparison 
purposes.  All energies
are in MeV.
}
\begin{tabular}{|c|lll|}
\hline
Theory   &  r.m.s & $a_v(c)$ & $ a_s(c)$ \\
& residual & & \\
\hline
SLy4 \cite{ch98} & 1.75  & -16.06 & 32.0 \\
\hline
SkP-based \cite{do84} & 1.75  & -16.11 & 31.1 \\
\hline
BSk4-based \cite{go03} & 1.65  & -16.03 & 29.6 \\
\hline
Skxce-based \cite{brown} & 1.55 & -16.10 & 31.0 \\
\hline
LD  & 3.1 & -15.6 & 23.3 \\
\hline
\end{tabular}
%  programs and files:
%     python process_skyr.py /scratch/refit/skp/skp_results_noq > 
%                            /scratch/refit/skp/skp_results_noq.dat
%     cd to directory   /scratch/refit/python/ditribute
%     python cheby.py /scratch/refit/skp/skp_results_noq.dat > cheby_skp_results.out  
%  for a_v and a_s, first get the refit Skyrme parameters using
%  python refit2.py /scratch/refit/brown/brown_results_noq
%  then  /scratch/refit/nuclear_matter/nuclear_matter.f90
%  xt data tables:  xt_*.dat  *= sly4,skp_refit,brown,goriely
%  for liquid drop:  /scratch/refit/python/lqs_ld_ee.py ../binding_tables/exp.tab
\end{table}

There may be multiple local minima in the parameter space of the
Skyrme functional, and other parameterizations in the literature
may reside in other minima. It is therefore of interest to see
what the refitting procedure produces for them. We have carried
this out for the SkP parameterization of ref. \cite{do84}, the
BSky4 of ref. \cite{go03} and the Skxce of ref. \cite{brown}. 
However, there is an important caveat in interpreting the
results. The quoted parameterizations were generated taking
different approximations for various non-Skyrme energy terms,
while our calculations here only vary the Skyrme parameters
themselves, keeping the same treatment of the other terms the
same as in the SLy4 calculation.  Thus, our extracted r.m.s. residuals
will not be directly comparable to the quoted residuals from the
original fits.  In Sect. IV below, we will
explicitly examine the effect of some of these ancillary
approximations on the fit.  Another caveat is that we start from the same
deformations as with SLy4.  The deformation is allowed to change
in the solution of the mean-field equations, but there could be
a lower energy state in some other well of the deformation
energy landscape.

Having obtained the wave functions for the different parameterizations, 
we extract the density integrals and eigenvectors and then apply the $N=4$ 
linear refit.  The results are  shown in Table II.  The main
eigenvectors of the SkP and the BSky4 were very similar to that of SLy4, and
the linear refits did not need a substantial adjustment of the parameters. 
The SkP has the same density dependence 
($\alpha = 1/6$) 
as the SLy4, and in fact we see from the table
that the quality of the fit is virtually identical.  The BSky4
has a density dependence  $\alpha= 1/3$, the value that is found
for the many-body theory of a dilute Fermi gas. 
Finally, we have also considered a parameterization with a density
dependence $\alpha=1/2$, the Skxce of ref. \cite{brown}.  In this
case, the original parameter set did not give an acceptable fit for
applying the linear refit.  We therefore made some iterations on
the fit to get a good starting point.

Comparing the
different parameter sets, we see that there is very little difference
between the qualities of the fits, all of them being in the range of
1.5-1.7 MeV.  
It is interesting to calibrate this number 
by comparing with the r.m.s. residual of the liquid drop model.
The result of fitting eq. (1) to the 2003 nuclear mass table \cite{au03} gives an
r.m.s. residual of 2.95 MeV; we quote in the last line of 
Table II the fit to even-even nuclei only.  
We see that the SCMF achieves a factor of two improvement 
in the calculated binding energies.  Of course the SCMF has twice as
many parameters, but as we just saw that many are superfluous from the
point of view of the binding energies.  Still, one might have hoped for
a more dramatic improvement given the computational cost of the SCMF
as compared to the liquid drop formula.

\section{Minimax fits}  

We now consider a completely different fitting criterion, the
Chebyshev norm.  The Chebyshev norm $\epsilon$ is 
defined as the maximum absolute value of the residuals $r_A =
E_{data} -E_{theory}$,
$$ 
\epsilon = \max_A |r_A|,
$$
We shall call this value the ``C-norm" for short.
The object of the fit is of course to minimize $\epsilon$, hence
the designation ``minimax fit".  
In general, if the theory has $N$ adjustable 
parameters, there will be $N+1$ members of the set that have a 
residual equal to $\epsilon$.  We call these the critical 
cases.  In searching for a better
theory, one can screen candidates by just testing them on
this set.  If the new theory does not produce a smaller
$\epsilon$ on the critical set, it can be immediately rejected.

We perform the minimax fit using the Chebyshev
norm as follows.  For an $N$-parameter theory, one first
selects a set of $N+1$ nuclei and 
makes the fit with them.  This can be done by the least squares method,
which yields equal residuals 
for $N+1$ nuclei.
Then the set is updated by replacing members with
other nuclei until a set is found that satisfies
the minimax condition.  It is easy to choose a nucleus to add to
the set--simply take the nucleus with the largest residual.  It
is not obvious which nucleus should be replaced.  The ascent 
algorithm described in ref. \cite{ch82} gives a procedure which we found
to be quite robust, usually coming to the critical set after less than
ten iterations.

\begin{table}
\caption{Liquid drop model, comparing least squares fits with minimax fits
of the 2003 and 1995 mass tables \cite{au03}.  The fit does not include light nuclei
($N$ or $Z < 8$).
}
\begin{tabular}{|c|c|c|l|}
\hline
Data set   &  r.m.s.& C-norm  & overbound critical nuclei \\
& (MeV)&(MeV)  & underbound critical nuclei\\
\hline
2003 & 2.9 & 9.2 & $^{40}$Ar, $^{76}$Se,$^{77}$Br,$^{229}$Fr \\
     &     &     & $^{100}$Sn,$^{132}$Sn \\
\hline
1995 & 3.0  &  8.0 & $^{73}$Ge,$^{101}$Nb,$^{230}$Ra \\
  &   & & $^{23}$O,$^{132}$Sn,$^{207}$Pb\\
\hline
2003 & 2.8  &  8.4  & $^{40}$Ar,$^{73}$As,$^{76}$Se,$^{229}$Fr \\
$\sigma<0.2$ Mev  &   & & $^{102}$Sn,$^{132}$Sn\\
\hline
\end{tabular}
% r.m.s. energies using /scratch/refit/python/lsq_ld.py and 
% /scratch/refit/binding_tables/exp.tab,exp.200.tab,exp1995.200.tab
% Chebyshev is done with /scratch/refit/python/Demo.py
\end{table}

We first apply the minimax fit to the 5-parameter
liquid drop model, eq. (\ref{ld}).  
Results are given in Table III.
For the first row, the
formula was fitted to the 2149 nuclei in the 2003 mass table 
having $N,Z\ge 8$.
Fitting to optimize the C-norm gives $\epsilon = 9.2$  MeV, while fitting
by least squares  gives an r.m.s. residual of 2.9 MeV \cite{note}.

The 6 critical nuclei of the minimax fit are given in the fourth column of
the table and their positions on
the chart of nuclides is shown in Fig.~4. We see that the critical 
nuclei are spread out over the entire mass range.
Two of the nuclei, $^{229}$Fr and $^{132}$Sn, are at the neutron-rich border of the mass table, 
and two are doubly magic: $^{100}$Sn and $^{132}$Sn.  One might expect that
the $N=Z$ line is problematic and indeed $^{100}$Sn is on the line and
$^{40}$Ar is near it.  The remaining nuclei are $^{76}$Se and $^{77}$Br
which do not have any particular properties that we are aware of.
\begin{figure}
\includegraphics [width = 13cm]{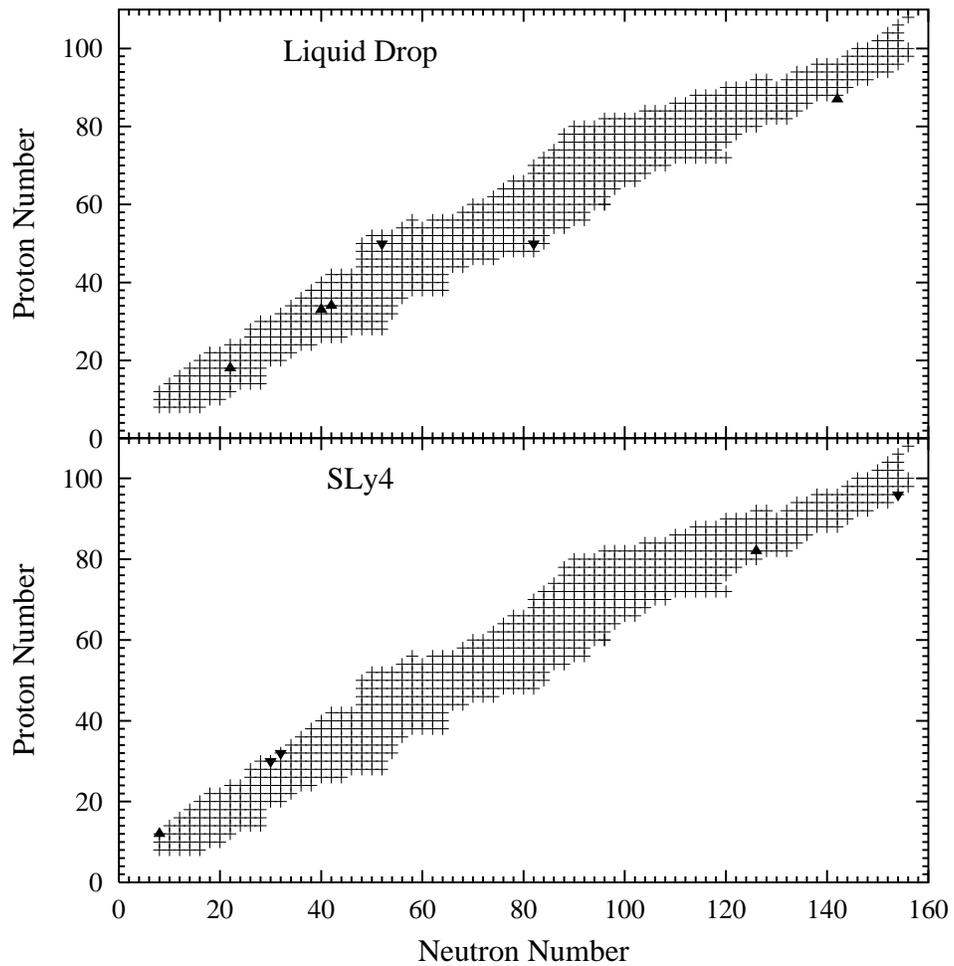}
\caption{Chart of nuclei showing the critical nuclei of the liquid
drop model and of the linearized refit of the SCMF based on the
SLy4 interaction.  Critical nuclei are indicated by triangles, with
the orientation of the triangle distinguishing overbound ($\Delta$) and underbound
($\nabla$) nuclei. The cross area shows the nuclei whose masses have been
measured.
}
\label{chart}
\end{figure}

It is interesting to compare with the fit to the 1995 mass table.
This is shown on the second line in the table.  The r.m.s. residual is 
actually lowered by the added nuclei in the 2003 table, but the C-norm is increased by 
more than
an MeV.  This suggests that the C-norm provides a better 
discrimination power to exhibit the increased
demands imposed by additional data.  Most of the added nuclei were
evidently easy ones for the formula, decreasing the least squares norm.
It is instructive to see which nuclei caused the changes to the
minimax fit. Several of the critical nuclei are different in the two fits, but
only one of these nuclei  is new to the 2003 mass table, namely $^{100}$Sn.
In fact this nucleus is not well determined
experimentally; the error in its quoted experimental mass is 0.7 MeV, larger
than the intended accuracy of the theory.    Restricting the 2003 mass
table to nuclei having errors less than 200 keV, the C-norm decreases by
0.8 MeV. This is shown on the third line of the Table. One sees that the nucleus
$^{100}$Sn is replaced by $^{102}$Sn and the C-norm
now is not so much greater than in 1995 mass table.
Thus, the minimax fitting process identifies $^{100}$Sn as an important
nucleus to measure more accurately, and it also provides justification
for the concentration of theoretical effort on the Sn isotope
chain.

We next discuss minimax fits of the Skyrme energy functional. 
Table IV shows the results of the $N=4$ refits to the Skyrme
energy functionals discussed earlier.  For SLy4, shown on 
the top line, the five critical nuclei include
two on the $N=Z$ line ($^{60}$Zn and $^{64}$Ge), one on the very
proton-rich side ($^{20}$Mg), and two heavy nuclei on the
neutron-rich side. One of them ($^{208}$Pb) is 
doubly magic and another is close to the heaviest known 
($^{254}$Cf).   
Overall, the nuclei are well spread as a function of mass number $A$.  
The SkP functional is similar to SLy4 in its density dependence and isospin
dependence of the spin-orbit interaction.  As may be seen in the Table,
the C-norm of our SkP-based functional is 
nearly identical to that of SLy4, but the set of critical
nuclei is quite different.  The set includes a nucleus on the $N=Z$ line, 
but the mass number is four units lower than the critical nucleus
$^{60}$Zn of SLy4.  The set lacks a doubly magic nucleus, and the only
magic number appearing is the $Z=82$ of the critical nucleus $^{214}$Pb.
The other underbound nuclei are $^{190}$W and $^{192}$Os, both of
which are deformed and on the border between prolate and oblate
shapes.  These results shown that the specific nuclei on the critical list
can change easily from one functional to another.  However, the value of the
C-norm is quite robust under small changes in the starting parameters.

We next consider a Skyrme parameterization based on the BSk4
functional of Goriely, et al. \cite{go03}.  Here the 
the density dependence is $\alpha=1/3$ rather than 1/6.
As mentioned earlier,
our results will not be directly comparable to theirs because of different
treatments of the pairing interaction and also additional terms
that they have added for their published fits.  Also, we have taken
the deformations from the SLy4 results, which could introduce some
error.  The results displayed on the 
third line in Table V show a slight improvement of
both norms with respect to the SLy4 interaction.  
The critical nuclei are different except for $^{20}$Mg and \pb.  There is also
an $N=Z$ nuclei in this critical set, $^{76}$Sr.  In fact, the
authors of ref. \cite{go03} introduce a phenomenological term
in the binding energies just to improve the fit near $N=Z$.  

The last functional we consider is based on the Skxce parameterization
having a density dependence $\alpha=1/2$ and a different isospin dependence
of the spin-orbit interaction.  As seen in Table V, it
yields a somewhat better C-norm than the $\alpha=1/6$ parameterizations,
which is in line with its better performance on the least squares norm.
Most of the critical nuclei are different from the previous cases,
with only $^{20}$Mg having appeared earlier.  Like the SLy4, there
are two midmass nuclei on the $N=Z$ line, but their identities are
different.  The only appearance of a spin-orbit shell closure in this
set is the $N=126$ neutron shell of the nucleus $^{218}$U.  Also, quite
different from the other fits, the most neutron-rich representative
is a midmass nucleus. 

\begin{table}
\label{skyrme-mm}
\caption{Minimax refits of Skyrme parameterizations, starting from parameter
sets from the literature, and experimental data from the 2003 mass
table, including only energies with errors less than 200 keV.
The critical nuclei are identified by 
proton number $Z$ and neutron number $N$ as $(Z,N)$.  The corresponding
fit with the liquid drop model is shown on the last line. 
}
\begin{tabular}{|c|c|c|c|}
\hline
Theory     & C-norm   & \multicolumn{2}{c|}  {critical nuclei} \\
\cline{3-4} 
& (MeV)  & overbound  & underbound \\
\hline
SLy4 \cite{ch98} & 4.8 & (12,8),(82,126) & (30,30),(32,32),(96,154) \\
\hline
SkP-based \cite{do84} &4.8  & (42,58),(82,132) & (28,28),(74,116),(76,116)  \\
\hline
BSk4-based \cite{go03}  & 4.7 & (12,8),(50,52),(82,128) & (38,38),(38,62)  \\
\hline
Skxce-based \cite{brown}  & 4.4 & (12,8),(92,126) & (34,34),(38,38),(38,64)  \\
\hline
LD  & 8.1 & (50,52),(50,132),(82,126) &(34,42),(88,142)  \\
\hline
\end{tabular}
% cd /scratch/refit/python/distribute
% python cheby.py  /scratch/refit/sly4/sly4_results.dat exp.200.tab >cheby_sly4_results_200.out
\end{table}

\section{Testing Theory Extensions} 
   There are two ways one can try to improve the theory.  One
is to introduce new theoretical ingredients or different approximations 
keeping the same set of parameters to be fitted, and the other is
to add terms
with additional parameters.  In this section we consider some examples
of the first approach, examining the effects on the C-norm.  

For our
first example, we ask how the treatment of the correlation energy
associated with center-of-mass motion affects the fit.  In 
principle SCMF is lacking that energy because the state it describes
is localized in space.  However, all the calculations done with 
the code {\tt ev8} include a one-body contribution to the center-of-mass
energy calculated by replacing the nucleon mass by its reduced mass.
A better approximation requires calculating the two-body
contributions to the center-of-mass energy; an approximation formula
for the total is given in ref. \cite{brown} as
\be
E_{cm} = -{3 \over 4} \left({45 \over A^{1/3}} - {25 \over A^{2/3} }
\right).
\ee
In Table V 
we apply
that correction to the five nuclei of the SLy4 critical set, first
taking out the one-body contribution and then adding $E_{cm}$
from eq. (3) to the
energies.  One sees only a tiny change in the resulting C-norm.
This shows that it is not worthwhile to put a lot of effort in 
making a better treatment of that term, when the goal is a
global improvement of the calculated binding energies.
Of course, this example was so simple that the correction
could have been easy applied to the mass table as a whole, but it
illustrates the point that one may be able to assess the effect of
the correction with a much smaller set.  

The next example we consider
has to do with the treatment the Coulomb energy.  In the SCMF calculations
reported above, the direct part of Coulomb energy was calculated
by solving the Laplace equation for the Coulomb potential using the
SCMF charge density.  The Coulomb exchange energy was calculated in
the local density approximation, as in eq. (3) of ref. \cite{brown}.
This underestimates the actual displacement energies
of isospin partners, as is well known as the Nolen-Schiffer anomaly.
We in fact see evidence of that in the BSk4 critical set 
which has a member $^{40}$Ti, an isospin partner of the stable
nucleus $^{40}$Ar. In fact the latter nucleus is critical for the
liquid drop model.  The experimental energy difference between 
these two nuclei is 29.3 MeV.  The theoretical difference in the
SLy4 theory is 27.5 MeV including the exchange, but 29.2 without
exchange.  This suggests eliminating the exchange term in doing
the refit.  Again, we try it on the critical nuclei first.  The
results are shown on the third line of Table \ref{improve}.  Here 
the fit is actually worse.  In this case we can identify the reason.  The
baseline theory has a serious problem in the overbinding of the
doubly magic nucleus $^{208}$Pb.  In the fit, the heavier nonmagic
nucleus $^{254}$Cf becomes critical with the opposite sign on the residual.
The parameter variation in the fits hardly affects this difference,
so any change in the theory that increases it will be difficult
to compensate for.  This is the case for the Coulomb exchange,
which is larger for $^{254}$Cf than for $^{208}$Pb.

The last 
example is inspired by the study of Bender, et al., \cite{be04} on the 
correlation energy.   A global table of correlation energies was calculated
in that work, with
the intention of adding them to the Skyrme energies.  Of course
the Skyrme parameters must be
refitted when the correlation energies are added.  We can take their
numbers for the critical nuclei to assess the quality of the improvement
(if any).  We consider
as a candidate improvement the simplest version of the correlation energy,
the effect of projection on angular momentum zero.
These projected energies are calculated with the {\tt promesse} code
\cite{be03b}.  
The difference
in energies from the two codes is taken as an additive correction,
and a refit is performed on the critical nuclei.
The results of
this exercise is shown on the fourth line of Table \ref{improve}.  
In this case the C-norm is decreased by 10\%.
This gives one some encouragement to carry out the program of 
ref. \cite{be04}, applying the projections to the
mass table as a whole.  Of course, it might still turn out that
other nuclei might become worse in the fit with no gain in 
the C-norm, but the correction at least passes a preliminary
screen.  The code {\tt promesse} also projects states of fixed
particle numbers $N,Z$ from the BCS wave function.  The effect
of the particle number projection can be isolated by turning
it off in the {\tt promesse} code, and using the difference
as a correction factor.  The result, shown on the last line,
is another 10 \% improvement in the C-norm.   We have also
checked that the improvements are additive taking the $J=0$
projection and the $N,Z$ particle number projections together.
\begin{table}
\label{improve}
\caption{Testing possible improvements of theory, starting from
the SCMF with the SLy4 interaction.  The putative improvement is
calculated on the SLy4 critical nuclei shown in Table IV.
}
\begin{tabular}{|c|c|}
\hline
Theory     & C-norm    \\
& (MeV)  \\
\hline
Baseline SLy4 & 4.84   \\
\hline
c.m. correlation & 4.69  \\
\hline
no Coulomb exchange & 4.89   \\
\hline
$J=0$ projection & 4.36  \\
\hline
$N,Z$ projection & 4.40  \\
\hline
\end{tabular}
\end{table}

\section{Final remarks}

  Fitting nuclear binding energies with theories based on quantum
many-body theory is a challenging task, partly due to the large data
set that needs to be computed.  We believe that work in developing
such theories can be reduced by using the Chebyshev norm as we
have illustrated.  To encourage this effort, our computer program that
carries out the minimax fits is available on the Web \cite{gfbweb}.

An issue still remains of which set of critical nuclei is best for
testing new theoretical ideas.  Unfortunately, each Skyrme
parameterization points to a different set.  We believe that
there are clusters of nuclei that come out close to the C-norm
limit, and that it shouldn't be so significant which set is used,
as long as all the nuclei of some critical set are included.  For
definiteness, we propose using the SLy4 set of five nuclei, since
it has a variety of types with a clear example of a doubly magic
nucleus.  

  For either the r.m.s. or the C-norm, the results of our refitting answers  
a basic question about the accuracy of SCMF
as applied to nuclear binding energies.  Namely, how does the
quality of SCMF fits including single-particle quantum 
mechanics compare with the liquid drop model which has no
quantum mechanics at all?  Comparing on the even-even nuclei
in the 2003 mass table, we found that the SCMF can double 
the accuracy for the r.m.s. norm and do somewhat better for
the C-norm.  The nominal number of Skyrme parameters is
more than twice the number of liquid drop parameters, but in fact all but about four 
could have been chosen from other considerations.  It is of
course reassuring that the SCMF is a more predictive theory,
and now the challenge is find good ways to extend it.

\begin{acknowledgments}
We thank H. Flocard, R. Furnstahl, D. Lunney and G. Audi for conversations 
and P.-H.~Heenen and M.~Bender for 
helpful advice and for making their codes available to us.
We also thank T. Duguet for a careful reading of the manuscript
which led to some changes and clarifications.  This work was 
supported by the U.S. Department of Energy,
Office of Nuclear Physics, under Contract
DE-FG02-00-ER41132.
\end{acknowledgments}


\begin{thebibliography}{99}

\bibitem{ch82} E.W. Cheney, ``Approximation theory," (American
Mathematical Society, Providence, 1982), p. 28-56.
\bibitem{ja83} F. James, Nucl. Inst. Meth. Phys. Res. {\bf 211} 145 (1983).
\bibitem{audi2} C. Borcea and G. Audi, "New methods for extrapolating
masse far from stability",
{\tt http://csnwww.in2p3.fr/AMDC/extrapolations/bernex.pdf}.
\bibitem{po95} B. Povh, K. Rith, C. Scholz and F. Zetsche, 
``Particles and Nuclei," (Springer, Heidelberg, 1995), p. 19.
\bibitem{be03}
  M. Bender, P.-H. Heenen, and P.-G. Reinhard,
  Rev. Mod. Phys. \textbf{75}, 121 (2003).
  %review on mf and density functional methods
\bibitem{be04} M. Bender, G.F.~Bertsch, and P.-H.~Heenen, nucl-th/0410023.
\bibitem{FH} R.P. Feynman, Phys. Rev. 56 340 (1939); 
H. Hellmann, Einfuehrung in die Quantenchemie (Deuticke, Leipzig, 1937).
\bibitem{ch98}
E. Chabanat, P. Bonche, P. Haensel, J. Meyer, and R. Schaeffer,
Nucl. Phys. \textbf{A635}, 231    (1998),   
Nucl. Phys. \textbf{A643}, 441(E) (1998).   
\bibitem{bo85} P. Bonche, et al., Nucl. Phys. {\bf A443} 39 (1985).
\bibitem{au03}
G. Audi, A. H. Wapstra, and C. Thibault,
Nucl. Phys. \textbf{A729}, 337 (2003);  
the data file is available at
{\tt http://www.nndc.bnl.gov/amdc/masstables/\\
Ame2003/mass.mas03}.
\bibitem{do84} J. Dobaczewski, et al., Nucl. Phys. {\bf A422} 103 (1984).
\bibitem{go03} S. Goriely, et al., Phys. Rev. C {\bf 68} 054325 (2003).
\bibitem{brown}  B.A. Brown, Phys. Rev. C {\bf 58} 220 (1998).
\bibitem{note}  The parameters
of the two fits are significantly different; the C-norm of 
the r.m.s. fit is 50\% higher than in the minimax fit and the 
r.m.s. of the residuals of the minimax
fit is 60\% larger than in the least squares fit.   
\bibitem{be03b} M. Bender and P.-H.~Heenen, Nucl. Phys. {\bf A713} 390
(2003); A. Valor, P.-H. Heenen, and P. Bonche, Nucl.
Phys. {\bf A671} 145 (2000).
\bibitem{gfbweb}  The computer program {\tt cheby.py} may be downloaded
from\\
{\tt http://gene.phys.washington.edu/\~{}bertsch/dist2.tar}.
\end{thebibliography}
\end{document}